\documentclass[11pt,a4paper]{article}

 \newcommand{\be}{\begin{equation}}
 \newcommand{\ee}{\end{equation}}
 \newcommand{\ba}{\begin{eqnarray}}
 \newcommand{\ea}{\end{eqnarray}}
 \newcommand{\bl}{\begin{equation}\begin{array}{ll}}
 \newcommand{\el}{\end{array}\end{equation}}
 \newcommand{\bll}{\begin{equation}\begin{array}{lll}}
 \newcommand{\bdm}{\begin{displaymath}}
 \newcommand{\edm}{\end{displaymath}}
 \def\bea{\begin{eqnarray}}
 \def\eea{\end{eqnarray}}
 \def\barr{\begin{array}}
 \def\earr{\end{array}}
\def\p{\partial}
\def\d{\partial}
\def\dif{\partial}

\def\f{\varphi}

\def\ep{\epsilon}
 \def\De{\Delta}

\def\half{\frac{1}{2}}

\def\lim{\rightarrow}

\def\hep{\hat{\epsilon}}

\def\ha{\hat{a}}

\def\hA{\hat{A}}
\def\hv{\hat{v}}

\def\hO{\hat{O}}

\def\tA{\tilde{A}}

\begin{document}
\raggedbottom

\title{{\bf Two-Dimensional Dilaton  Gravity and \\
 Toda - Liouville Integrable Models}}

\author{V. de Alfaro \thanks{vda@to.infn.it}\\{\small \it $^*$ Dip. Fisica
    Teorica, INFN, Accademia Scienze; v.Giuria 1, 10125 Torino IT}\\
A.T.~Filippov \thanks{Alexandre.Filippov@jinr.ru}~ \\
{\small \it {$^+$ Joint Institute for Nuclear Research, Dubna, Moscow
Region RU-141980} }}

\maketitle

\begin{abstract}

General properties  of a class of two-dimensional dilaton gravity (DG)
theories with multi-exponential potentials are studied and a subclass of
these theories, in which the equations of motion reduce to Toda and
Liouville equations, is treated in detail. A combination of parameters of
the equations should satisfy a certain constraint that is identified and
solved for the general multi-exponential model. From the constraint it
follows that in DG theories the integrable Toda equations, generally,
cannot appear without accompanying Liouville equations.
   We also show how the wave-like solutions of the general
Toda-Liouville systems can be simply derived. In the dilaton gravity
theory, these solutions describe nonlinear waves coupled to gravity as
well as static states and cosmologies.  A special attention is paid to
making the analytic structure of the solutions of the Toda equations as
simple and transparent as possible, with the aim to gain a better
understanding of realistic theories reduced to dimensions 1+1
   and 1+0 or 0+1.

\end{abstract}

\section{Introduction}

The theories of $(1+1)-$dimensional dilaton gravity coupled to scalar
matter fields are known to be reliable models for some aspects of
higher-dimensional black holes, cosmological models and waves. The
connection between higher and lower dimensions was demonstrated in
different contexts of gravity and string theory and, in several cases, has
allowed finding the general solution or special classes of solutions in
high-dimensional theories \footnote{See, e.g.,  \cite{BZ}-\cite{Venezia}
for a more detailed discussion of this connection, references, and
solution of some integrable two-dimensional and one-dimensional models of
dilaton gravity. }. A generic example is the spherically symmetric gravity
coupled to Abelian gauge fields and scalar matter fields. It exactly
reduces to a (1+1)-dimensional dilaton gravity and can be explicitly
solved if the scalar fields are constants  independent of
  the coordinates\footnote{This is not possible for arbitrary dependence
  of the potentials on the scalar fields, as will be clear in a moment.}.
  These solutions can describe interesting physical objects --
spherical static black holes and simplest cosmologies. However, when the
scalar matter fields, which presumably play a significant cosmological
role, are nontrivial, not many exact analytical solutions of
high-dimensional theories are known\footnote{See, e.g.,
         \cite{CGHS}, \cite{NKS}, \cite{ATF4}, \cite{VDA1}-\cite{ATF5};
         a review and further references can be found in \cite{Strobl},
  \cite{Kummer} and \cite{ATF5}.}.
  Correspondingly, the two-dimensional models of DG that
  nontrivially couple to scalar matter are usually not integrable.

  To construct integrable models of this sort one usually must make
 serious approximations, in other words, deform the original
 two-dimensional model obtained by direct dimensional reductions of
  realistic higher-dimensio\-nal theories.
 Nevertheless, the deformed models can qualitatively describe certain
 physically interesting solutions of higher-dimensional gravity or
 supergravity theories related to the low-energy limit of superstring
  theories.
 We note that several
 important four-dimensional space-times with symmetries defined by two
 commuting Killing vectors may also be described by two-dimensional
 models of dilaton gravity coupled to scalar matter. For example,
  cylindrical gravitational waves can be described by a $(1+1)-$dimen\-sio\-nal
  dilaton gravity   coupled to one scalar field
  \cite{Einstein}-\cite{Chandra1}, \cite{ATF3}.
 The stationary axially symmetric pure gravity (\cite{Ernst}, \cite{NKS})
  is equivalent to a $(0+2)-$dimensional dilaton gravity coupled to one scalar
field. Similar but more general dilaton gravity models were also obtained
in string theory. Some of them can be solved by using modern mathematical
methods developed in the soliton theory (see e.g. \cite{BZ},
\cite{Maison}, \cite{NKS}, \cite{Alekseev}).
 Note also that the theories in
 dimension 1+0 (cosmologies) and 0+1 (static states, in particular
 black holes) may be integrable in spite of the fact that their 1+1
 dimensional `parent' theory is not integrable without a deformation (see
 [23] and an example given in this paper).

  In our previous work (see, e.g., \cite{ATF2} - \cite{ATF5} and references
 therein) we constructed and studied some explicitly integrable models
 based on the Liouville equation.
 Recently, we attempted to find solutions of some realistic
     two-dimensional dilaton gravity models (derived from higher-dimensional
      gravity theories by dimensional reduction) using a generalized
 separation of variables introduced in \cite{W1}, \cite{ATF3}.
 These attempts showed that seemingly
natural ansatzes for the structure of the separation, which proved a
  success in previously studied integrable models, do not give
interesting enough solutions (`zero' approximation of a perturbation
theory) in realistic nonintegrable models. Thus an investigation of more
complex dilaton gravity models, which are based on the two dimensional
 Toda  chains, was initiated in \cite{ATF6}.

 At first sight
 it seems that it should be not difficult to find a potential in DG
 theory that will give integrable Toda equations of motion. However in
 reality it is not as simple as that, and the Toda theory may only
 emerge in company with a Liouville theory (this was mentioned in
 footnote in ref.~\cite{ATF6}). In fact, even the $N-$Liouville theory
 satisfies the same constraint.  It was known to the authors of \cite{ATF5}
  and \cite{ATF6} since long time
  but the meaning of this fact was not clearly understood.

In this paper we first introduce the general {\bf multi-exponential} DG
and present the equations of motion in a form that resembles the Toda
equations. In addition to the equations, in the DG theory one should
satisfy two extra equations which in General Relativity are called the
energy and momentum constraints. In the $N-$Liouville theory these
  constraints were explicitly solved but in the general case
  solving the constraints is a difficult problem which we
 discuss in Section~4.

Section~3 is devoted to the problem of reconstructing the dilaton gravity
from the `one-exponen\-tial' form of the equation of motion
\be
\d_u\d_v \,x_m\,=\,g_m\exp{\sum_n\,A_{mn}\,x_n}\,.
 \ee
This amounts to finding the matrix $\ha$ satisfying the matrix
equation\footnote{We call it the A-equation.} $\ha^T \hep\ha=\hA$ ($\hep$
is a diagonal matrix to be introduced later). Evidently, this equation may
have many solutions for a fixed matrix $\hA$ (e.g., if $\ha$ is a
solution, then $\hO\ha$, where $\hO^T\hep\hO=1$, is also a solution). The
  important fact is however that {\bf the solution is not possible for an
  arbitrary} symmetric matrix $\hA^T=\hA$. In Section~3 we establish the
class of `solvable' matrices $\hA$ (satisfying the A-condition) and
introduce a recursive procedure in order to find all possible solutions
 for any matrix satisfying the $A-$condition.

 We show that the Cartan matrices for simple Lie groups do not satisfy the
 A-condition and thus {\bf the generic DG cannot be reduced to the Toda equations}.
However, adding at least one Liouville equation to the Toda system (Toda -
Liouville System, or TL) solves this constraint  and in Section~4 we
briefly introduce the simplest form of solution of TLS in the case of the
$A_n$ Cartan matrices. We also discuss the problem of the energy and
 momentum constraints and solve the constraints for
 a class of Toda-Liouville theories.

 Finally,` we briefly discuss possible applications  of our results
 to the theory of black holes, cosmological models and waves
 which, at least in integrable theories, are closely related.

 \section{Multi - exponential model of (1+1)-dimensional dilaton\\ gravity
  minimally coupled to scalar matter fields.}

The effective Lagrangian of the (1+1)-dimensional dilaton gravity coupled
to scalar fields $\psi_n$ obtainable by dimensional reductions of a
higher-dimensional spherically symmetric (super)gravity can usually be
(locally) transformed to the form:
\be
 {\cal L}^{(2)} = \sqrt{-g}\left[ \f R(g) + V(\f,\psi) +
 \sum_{m,n} Z_{mn}(\f,\psi)\, g^{ij} \, {\dif}_i \psi_m \, {\dif}_j \psi_n \right] \,
\label{7}
\ee
(see \cite{ATF2} - \cite{ATF5} for a detailed motivation and examples).
 In Eq.(\ref{7}), $g_{ij}(x^0,x^1)$ is the (1+1)-dimensional metric with signature
 (-1,1), $g \equiv {\rm det}(g_{ij})$, $R$ is the Ricci curvature of
 the two-dimensional space-time with the metric
\be
ds^2=g_{ij}\, dx^i \, dx^j \, , \,\,\,\,\,\, i,j = 0,1 \, .
\label{4}
\ee
The effective potentials $V$ and $Z_{mn}$ depend on the dilaton
 $\f(x^0,x^1)$ and on $N-2$ scalar fields $\psi_n(x^0,x^1)$
 (we note that the matrix $Z_{mn}$ should be negative definite
 to exclude the so called `phantom' fields).
 They may depend on other parameters characterizing the parent
higher-dimensional theory (e.g., on charges introduced in solving the
equations for the Abelian fields).
         Here we consider the `minimal'
 kinetic terms with  diagonal and constant $Z$-potentials,
 $Z_{mn}(\f, \psi) = \delta_{mn} Z_n$.
  This approximation
 excludes the important class of the sigma - model - like
 scalar matter discussed, e.g., in \cite{Venezia}; such models can be
 integrable if $V \equiv 0$ and $Z_{mn}(\f,\psi)$ satisfy certain rather
 stringent conditions.
 In (\ref{7}) we also used the Weyl transformation to eliminate
 the gradient term for the dilaton.
   To simplify derivations, we write the equations of motion in the
 light-cone metric,  $ds^2 = -4f(u, v) \, du \, dv$.
 Now, by first varying the Lagrangian in generic coordinates and then passing
 to the light-cone coordinates we obtain the equations of motion
  ($Z_n$ are constants!)
 \be
 \p_u \p_v \f+f\, V(\f,\psi) = 0,
 \label{F.15}
 \ee
  \be
  f \p_i ({{\p_i \f} / f }) \, = \sum Z_n \,(\p_i \psi_n)^2\, ,
\,\,\,\,\,\,\,\,\, i=u,v \, .
 \label{F.17}
 \ee
 \be
 2Z_n \,\p_u \p_v\, \psi_n + f\, V_{\psi_n}(\f,\psi)= 0 \, ,
 \label{F.16}
 \ee
 \be
 \p_u\p_v\ln |f| + f V_{\f}(\f,\psi) = 0 \, ,
 \label{F.18}
 \ee
 where $V_{\f} \equiv \p_{\f} V$, $V_{\psi_n} \equiv \p_{\psi_n} V$.
 These equations  are not independent. Actually, (\ref{F.18}) follows
from (\ref{F.15}) $-$  (\ref{F.16}). Alternatively, if  (\ref{F.15}),
(\ref{F.17}), and (\ref{F.18}) are satisfied, one of the equations
(\ref{F.16}) is also satisfied. Note that the equations may have the
solution with $\psi_n = \psi_n^{(0)} = \textrm{const}$ only if
 $V_{\psi_n}(\f , \psi_n^{(0)}) \equiv 0$.

The higher-dimensional origin of the Lagrangian (\ref{7}) suggests that
the potential is the sum of exponentials of linear combinations of
 the scalar fields
and of the dilaton $\f$ \footnote{Actually, the potential $V$ usually
contains terms
       non exponentially depending on $\f$ (e.g., linear in $\f$),
       and then the exponentiation of $\f$ is only an approximation,
       see the discussion in \cite{ATF5}.}.
     In our previous work \cite{ATF5}
 we studied the constrained Liouville model, in which
the system of equations of motion (\ref{F.15}), (\ref{F.16}) and
(\ref{F.18}) is equivalent to the system of independent Liouville
equations for the
 linear combinations of  fields $q_n \equiv F + q_n^{(0)}$, where
 $F \equiv \ln|f|$. The easily derived solutions of these equations should
 satisfy  the constraints (\ref{F.17}), which was the most difficult part of
 the problem. The solution of the whole problem revealed an interesting
 structure of the moduli space of the solutions that allowed us to easily
 identify static, cosmological and wave-like solutions and effectively
 embed these essentially one-dimensional (in a broad sense) solutions
 into the set of all two-dimensional solutions and study their analytic
 and asymptotic properties.

 Here we propose a natural generalization of the Liouville model to the
 model in which the fields are described by the Toda equations (or by
 nonintegrable deformations of them). To demonstrate that the model shares
 many properties with the Liouville one and to simplify a transition from the
          integrable models to nonintegrable theories we suggest a different
 representation of the Toda solutions which is not directly related to
 their group - theoretical background.

 Consider the theory defined by the Lagrangian (\ref{7})
 with the potential ($Z_n =-1$):
 \be
 V = \sum_{n=1}^N 2g_n \exp{q_n^{(0)}} \, , \qquad
 q_n^{(0)} \equiv a_n \f + \sum_{m=3}^{N} \psi_m a_{mn} \, .
 \label{8}
 \ee
 In what follows we also use
 \be
 q_n \equiv  F + q_n^{(0)} \equiv \sum_{m=1}^{N} \psi_m a_{mn} \, ,
 \label{9aa}
 \ee
where $\psi_1 + \psi_2 \equiv \ln{|f|} \equiv F$, $\psi_1 - \psi_2 \equiv
\f$ and hence $a_{1n} = 1 + a_n$, $a_{2n} = 1 - a_n$.

 Rewriting the equations of motion in terms of $\psi_n$
 we find that  Eqs.~(\ref{F.15}) - (\ref{F.18}) are equivalent to
 $N$ equations of motion for $N$ functions $\psi_n$
 ($\varepsilon$ is the sign of the metric $f$),
\be
 \p_u \p_v \psi_n  =
  \varepsilon  \sum_{m=1}^{N} \epsilon_n a_{nm} g_m \exp{(q_m)} \quad
 ( \epsilon_1 = -1, \,\,\, \epsilon_n = +1 \,\, {\rm if} \,\, n \geq 2 \,)\, ,
\label{10}
\ee
and two constraints,
\be
 C_i \equiv \p_i^2 \f + \sum_{n=1}^N  \epsilon_n  (\p_i \psi_n)^2 =  0,
 \,\,\,\,\,\,\, i=u,v \, .
 \label{11}
\ee
 With arbitrary parameters $a_{nm}$, these equations of motion are not
integrable.
 But as proposed in \cite{A2} - \cite{ATF1}, \cite{ATF2} \cite{ATF5},
 Eqs.(\ref{10}) are integrable and the constraints (\ref{11}) can be solved
 if the $N$-component  vectors $v_n \equiv (a_{mn})$ are pseudo-orthogonal.

 Now, consider more general nondegenerate matrices $a_{mn}$ and
       define the new scalar fields $x_n$:
\be
       x_n  \equiv  \sum_{m=1}^{N} a_{nm}^{-1} \epsilon_m \psi_m  \, ,
   \,\,\,\,\,\,\,\,\,\,\,\,\,\,\,
       \psi_n  \equiv  \sum_{m=1}^{N} \epsilon_n a_{nm} x_m  \, .
 \label{12}
\ee
 In terms of these fields, Eqs.(\ref{10})  read as
\be
  \p_u \p_v x_m  \equiv  \varepsilon g_m
     \exp ({\sum_{k,n=1}^{N}  \epsilon_n a_{nm} a_{nk} x_k}\,)
     \equiv  \varepsilon g_m \exp ({\sum_{k=1}^{N} A_{mk} x_k}\,)  \, ,
 \label{13}
\ee
 and we see that the symmetric matrix
   \be
  \hA \,\equiv \,\ha^T \,\hep \,\ha \, , \qquad \ep_{mn} \equiv
  \ep_m \ \delta_{mn} \,,
\label{14}
    \ee
     defines the main properties of the model.

 If $\hA$ is a diagonal matrix  we return to the $N$-Liouville model.
 If $\hA$ were the Cartan matrix of
 a simple Lie algebra, the system (\ref{13})
 would coincide with the corresponding
 Toda system, which is integrable and can be more or less explicitly
         solved (see, e.g., \cite{Leznov}, \cite{Saveliev} ).
  However, it can be shown that the Cartan matrices
         of the simple Lie algebras (symmetrized when necessary)
          cannot be represented in the form (\ref{14}).
  Nevertheless, a very simple extension of the Toda equations obtained by adding one or
  more Liouville equations can solve this problem. In fact, a symmetric
  matrix $A_{mn}$ that is the direct sum of
  a diagonal $L\times L$-matrix $\gamma_n^{-1} \delta_{mn}$
  and of an arbitrary symmetric matrix $\bar{A}_{mn}$,
  can be represented in form (\ref{14}) if the sum of $\gamma_n^{-1}$
  is a certain function of the matrix elements $\bar{A}_{mn}$ .
  If $\bar{A}_{mn}$ is a Cartan matrix, the system (\ref{13})
  thus reduces to $L$ independent Liouville (Toda $A_1$) equations and the
  higher-rank Toda system (TLS).

 The solution of TLS can be derived in several ways. The most general one
 is provided by the group-theoretical construction described in
 \cite{Leznov}, \cite{Saveliev}. Here, in Section~4 we outline an
 analytical method directly applicable to solving $A_N$ TLS proposed in
 \cite{ATF6}. However, solving the equations of motion is not the whole
   story.
  Once the equations are solved, their solutions must be
 constrained to satisfy the zero energy-momentum conditions (\ref{11})
 that in terms of $x_n$ are:
 \be
 -C_i \, = \, 2\sum_{n=1}^N \ \p_i^2 x_n \, - \,\sum_{n,m =1}^N \ \p_i x_m \
 A_{mn} \, \p_i x_n \, = \,0 \,, \quad i=u,v \, .
 \label{15}
 \ee
 In the $N$-Liouville model the most difficult
 problem was to satisfy the constraints (\ref{15}) but this problem was
 eventually solved. In the general nonintegrable case
            of an arbitrary matrix
 $\hA$, we do not know even how to approach this problem.
 The Toda case is discussed below.

  To study the general properties of the solutions
 of equations (\ref{13}) and of the constraints (\ref{15})
 we first rewrite the general equations in a form that is particularly
 useful for the Toda-Liouville systems. Introducing notation
  \be
 X_n \equiv \exp (-\half A_{nn} x_n) \ , \,\,\,\,\,\,
  \Delta_2 (X) \equiv X\ \p_u \p_v X - \p_u X \ \p_v X , \,\,\,\,\,\,
  \alpha_{mn} \equiv -2 A_{mn} / A_{nn} \,,
 \label{16}
 \ee
 it is easy to rewrite Eqs.(\ref{13}) in the form:
 \be
 \Delta_2 (X_n) =
  -\half \varepsilon \ g_n A_{nn} \prod_{m \neq n} X_m^{\alpha_{nm}} \, .
 \label{17}
 \ee
 The multiplier $|-\half \varepsilon \ g_n A_{nn}|$ can be removed by using
 the transformation $x_n \mapsto x_n + \delta_n$ and the final (standard)
 form of the equations of motion is
 \be
 \Delta_2 (X_n) =
  \varepsilon_n  \prod_{m \neq n} X_m^{\alpha_{nm}} \, ,
  \qquad   \varepsilon_n \equiv \pm 1 .
 \label{18}
 \ee

 These equations are in general not integrable. However, when $A_{mn}$
     are Toda plus Liouville matrices, they simplify to integrable equations
 (see \cite{Leznov}). The Liouville part is diagonal while the Toda part
 is non-diagonal.
   For example, for the Cartan matrix of $A_N$, only the
   near-diagonal elements of the matrix $\alpha_{mn}$
   are nonvanishing, $\alpha_{n+1,n-1} = \alpha_{n-1,n+1} = 1$.
 This allows one to solve Eq.(\ref{18}) for any $N$.
 The parameters $\alpha_{mn}$ are invariant w.r.t.
 transformations $x_n \mapsto \lambda_n x_n + \delta_n$.
   This means that the non-symmetric Cartan matrices of
 $B_N$, $C_N$, $G_2$, and $F_4$ can be symmetrized while not
 changing the equations. In this sense,
 $\alpha_{mn}$ are the fundamental parameters of the equations of
 motion. From this point of view, the characteristic property of the
 Cartan matrices is the simplicity of Eqs.(\ref{18}) which allow one
 to solve them by a generalization of separation of variables.
 As is well known, when $A_{mn}$ is the Cartan matrix of any simple
  algebra, this procedure gives
 the exact general solution (see \cite{Leznov}). In  Section~4 we
 show how to construct the exact general solution for the $A_N$ Toda system
 and write a convenient representation for the general solution that
 differs from the standard one given in \cite{Leznov}.

 Unfortunately, as we emphasized above, solving equations (\ref{18})
 is not sufficient for finding the solution of the whole problem.
 We also must solve the constraints (\ref{15}), and this is a more
 difficult task. In our previous papers we succeeded in solving the
 constraints of the $N$-Liouville theory. So, let us try to formulate the
 problem of the constraints in the Toda-Liouville case as close as possible
 to the $N$-Liouville case.
    First, it is not difficult to show that
    $\d_v\,C_u\,=\,\d_u\,C_v\,=\,0$ and thus
    $C_u = C_u (u)$, $C_v = C_v (v)$ as in the Liouville case.
    To prove this one should differentiate (\ref{15}) and use (\ref{13})
    to get rid of $\p_u \p_v x_m$ and $\p_u \p_v x_n$.

    Up to now we considered an arbitrary symmetric matrix $\hA$.
 At this point we should use a more detailed information about $A_{mn}$ and
 about the structure of the solution.
  To see whether the constraints can be solved we first rewrite
   them in terms of $X_n$ and then consider
  the Toda - Liouville matrices and
  the explicit solutions of the equations.
 It is not difficult to see that the constraints (\ref{15}) can be written
 in the form ($i=u$ or $i=v$ and the prime denotes $\partial_i$):
  \be
  \label{f30}
  {1\over 4} C_i =
 \sum_{n=1}^N {1 \over A_n}{X_n^{''}\over X_n} \,+\, \sum_{m<n}^N
 {2A_{mn}\over A_mA_n}\, {X_m^{'}\over X_m}\,
 {X_n^{'}\over X_n}\, .
\ee
The first term looks exactly as in the case of the $N-$Liouville model.
However, in the Liouville case we also knew that
 \be
 \label{f35}
 \d_u\biggl(X_n^{-1} \,\d_v^2 X_n\biggr) = \,0,
 \quad \d_v\biggl(X_n^{-1} \,\d_u^2 X_n\biggr)=\,0,
 \ee
which is not true in the general case. Moreover, the first and the second
terms in r.h.s. of Eq.(\ref{f30}) are in general not functions of a single
variable (above we have only proved
  that in general $C_u=C_u(u)$ and $C_v=C_v(v)$.

Nevertheless, let us try to push the analogy with the Liouville case as
far as possible, at least in the integrable Toda - Liouville case. Thus,
suppose that the first $N_1$ equations are the Toda ones and the remaining
$N_2=N-N_1$ equations are the Liouville ones. This means that
$A_{mn}=\tA_{mn} $ ($1\leq m,n\leq N_1$), where $\tA_{mn}$ is a Cartan
matrix while for $N_1+1\leq m,n\leq N$ we have
  $A_{mn}=\delta_{mn} \gamma_n^{-1}$.
Then the constraints split into the Toda and the Liouville parts:
 \be
 \label{f40}
 {1\over 4} C_i = \sum_{n=1}^{N_1}\,{1\over A_n}\,{X_n^{''} \over X_n} \,+\,
  \sum_{m<n}^{N_1}\, {2A_{mn}\over A_mA_n}\, {X_m^{'}\over X_m}\,
  {X_n^{'}\over X_n}\, + \,\sum_{n=N_1+1}^N\, \gamma_n {X_n^{''} \over X_n}\,.
 \ee
  They are significantly different: first, because
  the Liouville solutions $X_n$ for $n\geq N_1+1$
 satisfy the second order differential equation while the Toda solutions
 $X_n$ satisfy higher order ones (see Section 4).
   In the general $A_N$ Toda case $X_1$ can be written as
 \be
 \label{f45}
  X_1=\sum_{i,j=1}^{N+1}\,a_i(u) \,b_i(v) \,,
   \ee
     while in the Liouville case the solution is simply the sum
     of two terms and (see Section 4).
     Moreover, for the Liouville solution we have
 \be
 \label{f55}
 X^{-1} \d_u^2 X = {a_1^{''}(u)\over a_1(u)}
 ={a_2^{''}(u)\over a_(u)} \,,
 \qquad
 X^{-1} \d_v^2 X = {b_1^{''}(v) \over b_1(v)}
 = {b_2^{''}(v) \over b_2(v)} \,,
  \ee
while in the Toda case everything is much more complex.

To understand better this fact we consider the case $N_1=2$, $N=3$ with
$A_{mn} (1\leq m,n\leq 2)$ being the $A_2-$ Cartan matrix and
$A_{3n}=\delta_{3n} A_3$.
 Using $A_1=A_2=2$, $A_{12}=A_{21}=-1$, we find
 \be
 \label{f60}
 {1\over 2} C_i\,=\, \biggl({X_1^{''} \over X_1} +{X_2^{''}\over X_2} -
 {X_1^{'}\over X_1}\cdot {X_2^{'}\over X_2}\biggr) \,-\,
  4{X_3^{''}\over X_3} \, = 0
 \ee
 where $X_2 = \varepsilon_1 \De_2(X_1)$, $\varepsilon_2 = \pm 1$,
 $X_3$ is the Liouville solution
 (note that according to the constraint on
$A_{ij}$ we have in this case $\gamma_3=A_3^{-1}=-2$). Although we know
that $X_3^{''}/X_3$ and $C_i$ are functions of one variable, we do not
have at the moment simple and explicit expressions for $C_i$.
 Indeed, using (\ref{f45}) it is not difficult to find that
 \be
 \label{f65}
  \d_v (X_1^{-1} \d_u^2 X_1) = \,
   = \, \biggl(\sum_{j=1}^3 a_j\,b_j \biggr)^{-2}\,\sum_{i>j}
  W^{'}[a_i,a_j]\, \,W[b_i,b_j] \,\not=\,0\,.
  \ee
 So, we should first write the explicit expression for $X_2(u,v)$ in
 terms of $a,\,b,$ and then derive the complete first term in $C_i$.
 We construct solutions of the $A_2 + A_1$ constraints in Section 4.

\section{Solving $\ha^T\,\hep\,\ha\,=\,\hA$}
 In this section we show how to solve Eq.(\ref{14}) for the matrix $\ha$ in
 the standard DG. This is possible if and only if $\hA$ satisfies certain
 conditions, which we explicitly derive.
  First, $\det \hat{A} = -\det \hat{a}^2<0$.
  This restricts the matrices $\hA$ of even order but is not so severe a
 restriction for the odd order matrices. In fact, we can then change sign
 of $\hA$ and of all the variables $x_n$ and the only effect will be
 that all $\varepsilon_n$ in Eq.(\ref{18}) change sign. If these signs are
 unimportant and the two systems of equations may be considered as
 equivalent, the restriction does not work. As the
 determinants of all (symmetrized) Cartan matrices for simple groups are
 positive (and their eigenvalues are positive), it follows
 that the even-order Cartan matrices do not satisfy this restriction.
 A more severe restriction is related to the special structure of
 the matrices $a_{mn}$ in (\ref{9aa}). In consequence, the matrix $\hA$
 must satisfy one equation that we derive and explicitly solve below.

 Let us now take the general $N \times N$ matrix $\ha$ of DG, with the only
 restriction: $a_{1n} = 1+a_n$ and $a_{2n} = 1-a_n$.
      The equations defining $a_{mn}$ in terms of $A_{mn}$ are
\be
\label{1}
-2(a_m+a_n) \,+V_m \cdot V_n \,=\, A_{mn} \,,
 \qquad    -4a_n\,=\,A_n-V_n^2\,,
 \qquad  m,n=1,...,N
\ee
where we introduced notation  $V_n\,\equiv\,(\,a_{3n},...,a_{Nn})$.
 As follows from (\ref{1}),
 our $N$ vectors $V_i$ in the $(N-2)-$dimensional space have $N(N-2)$
 components and satisfy  $N(N-1)/2$ equations:
 \be
 \label{1a}
 (V_m-V_n)^2 \,=\,A_m+A_n-2A_{mn}\,, \qquad m>n,\,\,\,m,n=1,...,N.
 \ee
These equations are invariant under $(N-2)\,(N-3)/2$ rotations of the
$(N-2)-$ dimensional space and under ${N-2}$ translations.
  It follows that the vectors $V_m$ in fact depend on
 \bdm
 N(N-2)\,- (N-2) - \half\,(N-2)\,(N-3) \,=\,{\half}\,(N-2)\,(N+1)
 \edm
 invariant parameters. The $N(N-1)/2$ equations should define
 $(N-2)\,(N+1)/2$ parameters.
 Thus one can see that the number of
   equations minus the number of parameters is equal to one,
   and thus one of the equations will give a relation between the parameters.

 It is possible to give a more constructive approach directly
  utilizing  the invariant equations that
  follow from the equations $(\ref{1a})$ above.
  Define $v_k\equiv V_k-V_1$, where
 $k=2, ...,N$. Then, from $(\ref{1a})$ we have:
\bdm
  v_k^2\,\equiv\,(V_k-V_1)^2 \,=\,A_1+A_k-2A_{1k} \,\equiv \,
  \tilde{A}_{1k}\, ,
\edm
  \bdm
   (v_k-v_l)^2 \,\equiv \,
   \tilde{A}_{1k} \,+\, \tilde{A}_{1l} \,-\, 2v_k \cdot v_l \, ,
   \qquad k>l; \,\,\,\, k,l=2,...,N\, .
  \edm
  Thus the general invariant equations for $v_k$ can be written:
 \be
 v_k \cdot v_l\,=\, A_1-A_{1k}-A_{1l}+A_{kl} \,, \qquad k \geq l \,.
 \label{2a}
 \ee
  As these equations are valid also for $l=k$ we have $N(N-1)/2$
  equations for the same number of the invariant parameters
  $v_k \cdot v_l$, as it should be.
  But, of course, there is one relation between these parameters because
  there exist a linear relation between $N-1$ vectors $v_k$ in the $(N-2)-$
  dimensional space. For example, $v_N^2$ can be expressed in terms of the
  remaining parameters $v_2^2, ..., v_{N-1}^2$ and $v_k \cdot v_l$, $k>l$
  (their number is $(N-2)(N+1)/2$, as above).
  As the equations for $v_k$ express $v_k \cdot v_l$ in terms of the matrix
  elements $A_{kl}$, we thus can derive
  the necessary relation between $A_{kl}$ (e.g. an expression
  of $A_1\equiv A_{11}$ in terms of the remaining matrix elements).

    Using the vectors $v_k$ we can give an explicit construction of the
 solutions and derive the constraint on the matrix elements $A_{mn}$.
 The construction of the solution of the equations for
 $a_{mn}$ can be given as follows. It is not difficult to understand
 that we only need to find the unit vectors,
 \be
 \hv_k \equiv {v_k / |v_k|}\,=\,v_k\,\tilde{A}_{1k}^{-1/2} \,,
 \ee
 in any fixed coordinate system in the $(N-2)-$ dimensional space. Then we
 can reconstruct the general solution by applying to $\hv_k$ rotations
 and translations (i.e. choosing arbitrary $a_{n1}$, $n=3,...,N$).
 Let us introduce the temporary notation
\be
 c_{kl}\equiv \cos\theta_{kl} \equiv
   \hv_k \cdot \hv_l \,=\,
 (A_1-A_{1k}-A_{1l}+A_{kl}) \, (\tilde{A}_{1k} \, \tilde{A}_{1l})^{-1/2}.
\ee
As  $v_k=(a_{3k}-a_{31}, ..., a_{Nk}-a_{n1}\,)$, we denote
$\alpha_{nk}\equiv (a_{nk}-a_{n1})/|v_k|$ and thus $\hv_k=(\alpha_{3k},
..., \alpha_{Nk})$. Choosing the coordinate system in which
 $\hv_2=(1,0,..0)$ we see that
 $\alpha_{3k}=c_{k2}\equiv \cos\theta_{2k}$ and $\hv_3$
 can be chosen with two nonvanishing components,
\be
 \hv_3 =(c_{23}, s_{23},0,...,0)\,,
\ee
 where $s_{23} \equiv \sin\theta_{23}$ and in general
 $s_{kl}=\sin \theta_{kl}$.
  The further invariant parameters $\alpha_{nk}$ can be derived recursively.
  The vectors $\hv_k,...,\hv_N$ for $k\geq 4$ are constructed as follows (it is
easy to check!). We take $\alpha_{3k}=c_{2k}$, $\alpha_{nk}=0$ if $k\leq
N-2$ and $n\geq k+2$.  Thus
\be
\hv_k=(c_{2k},\alpha_{4k}, \alpha_{5k}, ...,\alpha_{(k+1)k}, 0, 0...)
\ee
 and the parameters $\alpha_{nk}$ can be
 recursively derived from the relations ($k\geq 4$)
\be
\sum_{n=4}^{l+1}\,\alpha_{nk}\alpha_{nl}=c_{kl}-c_{k2}c_{l2}\,;
 \quad k>l , \qquad \qquad
\sum_{n=4}^{k+1} \,\alpha_{nk}^2  \,=\, s_{k2}^2 ,\quad k\leq N-1\,.
\ee
 The normalization condition for
$\hv_{N}$ (not included in the above equations),
\be
\sum_{n=4}^N\, \alpha_{nN}^2 \,=\, s_{N2}^2 \,,
\ee
 then gives  a relation
 between the $c_{kl}$'s  (and thus between the $A_{ij}$'s).

 Using this solution we can find the expression for $A_1\equiv A_{11}$ in
 terms of $A_{kl}$. However, this derivation is rather
 awkward. It can be somewhat simplified if we consider simpler matrices
 $A_{kl}$ for which $A_{1k}=A_{k1}=0$, $k \neq 1$. Then one can find that
 the equation for $A_1$ is linear and
 thus has the unique solution.
 Nevertheless it is not a good idea to derive the
 constraint on $A_{kl}$ in this rather indirect way. The linearity of the
 constraint in $A_1$ suggests that there exists a simple and general
 formula directly expressing $A_1$ in terms of the other elements $A_{kl}$.

 The simplest way to find $A_1$ in terms of the other $A_{ij}$ is the
following: one of the vectors $v_2,\,v_3,\, ...,\,v_N$ must be given by a
 linear combination of $N-2$ other vectors. Suppose that
 \be
 \label{c20}
v_2=\sum_{p=3}^N \, v_p\,z_p\,.
 \ee
 Then we can find $z_p$ in terms of
 $A_{mn}$ by solving the equations
 \be
 \label{4a}
 v_p \cdot v_2\,=\,\sum_{q=3}^N\, (v_p \cdot v_q)\,z_q\,,\quad p=3,...,N\,.
 \ee
 The solution is given by $z_p=D_p/D$, where $D$ is the determinant
 of the $(N-2)\times (N-2)$  matrix $(v_p \cdot v_q)$, and the $D_p$
 are the determinants of the same matrix but with the $p-$th column replaced
 by $(v_p \cdot v_2)$.

  Now it is clear that the expression of $v_2^2$ in terms of the solution
  of (\ref{4a}),
  \be
 \label{5a}
 v_2^2\,=\,\sum_{q=3}^N\, (v_2\cdot v_q) \, \,z_q\,
 =\,\sum_q\,(v_2 \cdot v_q) \,\cdot\,{D_q/ D}, \,
  \ee
 gives us the desired constraint on $A_{mn}$.
 Using (\ref{2a}) we rewrite it in the form
\be
\label{b10}
 (A_1+A_2-2A_{12})\,D\,=\,\sum_{p=3}^N\,
 (A_1+A_{p2}-A_{12}-A_{1p}\,)\, D_p\, ,
\ee
 where the determinants $D$ and $D_p$ should be expressed in terms of
  $A_{mn}$.  They evidently depend on $A_1$ linearly and thus
 Eq.(\ref{b10}) is at most quadratic in $A_1$.
 In fact, it is just linear. To prove this it is sufficient to show that
 \be
 \label{6a}
{dD \over dA_1}\,=\,\sum_{p=3}^N\, {dD_p \over dA_1}\,.
 \ee
 This is not very difficult but we omit the proof because of the space
 restrictions.

\section{Solution of the $A_N$ Toda system}
 The equations (\ref{18}) for the
 $A_N$-theory are extremely simple,
 \be
 \Delta_2 (X_n) =
  \varepsilon_n X_{n-1} X_{n+1} \, , \,\,\,\,\,\,\,\,
  X_0 \mapsto 1 \ , \,\,\,\, X_{N+1} \mapsto 1 , \,\,\,\,
 n = 1,...,N ,
 \label{19}
 \ee
 where $\varepsilon_n^2 =1$. As is well known, their solution can be
 reduced to solving just one higher-order equation for $X_1$
 by using the relation (see \cite{Leznov}):
 \be
 \Delta_2 (\Delta_n (X)) =
 \Delta_{n-1}(X) \ \Delta_{n+1}(X) \, , \,\,\,\,\,\,\,\,
 \Delta_1(X) \equiv X , \,\,\,\,\,\,  n \geq 2 \, .
 \label{20}
 \ee
 Indeed, using Eqs.(\ref{19}), (\ref{20}) one can prove that for $n \geq 2$
 \be
  X_n = \Delta_n (X_1) \prod_{k=1}^{[n/2]} \varepsilon_{n+1-2k} \, ,
 \label{20a}
 \ee
 where the square brackets denote the integer part of $n/2$.
 Thus the condition $X_{N+1} = 1$ gives the equation for $X_1$,
 \be
 \Delta_{N+1} (X_1) = \prod_{k=1}^{[(N+1)/2]} \varepsilon_{N+2-2k} \,
  \equiv  \tilde{\varepsilon}_{N+1} \,=\, \pm 1 \, .
 \label{21}
 \ee
 This equation looks horrible but it is known to be exactly soluble
 by a special separation of variables, Eq.(\ref{f45}).
 We present its solution in
 a form that is equivalent to the standard one \cite{Leznov} but is
 more compact and more suitable for constructing effectively
 one-dimensional solutions, generalizing those studied in \cite{ATF5}.

     Let us start with the Liouville ($A_1$ Toda) equation
     $\Delta_2(X) = \tilde{\varepsilon}_2 \equiv \varepsilon_1$
 (see \cite{DPP}, \cite{Gervais}, \cite{Leznov}, \cite{ATF5}).
    Calculating the derivatives of $\Delta_2(X)$ in the variables $u$ and $v$,
it is not difficult to prove Eqs.(\ref{f35}).
  It follows that there exist some `potentials' ${\cal{U}}(u)$,
${\cal{V}}(v)$ such that
\be
\label{23}
\p_u^2 X \,-\,{\cal U}(u) \, X\,=\,0 \, , \qquad \p_v^2 X \,-\,{\cal V}(v)
\, X\,=\,0 \,,
\ee
 and thus $X$ can be written in the `separated' form
 given in (\ref{f45}) with $N=1$
 where $a_i(u)$, $b_j(u)$ ($i, j = 1,2$) are linearly
 independent solutions of the equations (Eq.(\ref{f55})
 follows from this):
\be
\label{25}
a''_i(u) \, - \,{\cal U}(u) \, a_i(u) \, = \,0,
\qquad
b''_i(v)  \, - \, {\cal V}(v) \, b_i(v) \, = \, 0 \, .
\ee
 For $i=1$ these equations define the potentials for any choice of
  $a_1$, $b_1$, while
 $a_2$, $b_2$ then can be derived from the Wronskian first-order equations
\be
\label{26}
 W[a_1(u), a_2(u)] = w_a  \ , \,\,\,\,\,\,\,\,
 W[b_1(v), b_2(v)] = w_b \ , \,\,\,\,\,\,\,\,
     w_a w_b = \varepsilon_1 \ .
\ee

 We have repeated this well known derivation because it is
 applicable to the $A_N$ Toda equation (\ref{21}).
 By similar derivations it can be shown that
 $X_1$ satisfies the equations
 \be
 \label{27}
 \p_u^{N+1} X + \sum_{n=0}^{N-1} {\cal U}_n(u) \ \p_u^n X = 0 \, ,
 \qquad
 \p_v^{N+1} X + \sum_{n=0}^{N-1} {\cal V}_n(v) \ \p_v^n X = 0 \, .
 \ee
 Thus the solution of (\ref{21}) can be written in the
 same `separated' form (\ref{f45}), where now
 $a_i(u)$, $b_i(v)$ ($i = 1,...,N+1$) satisfy the ordinary linear differential
 equations corresponding to (\ref{27}),
 with the constant Wronskians normalized by the conditions
 (one can choose any other normalization in
 which the product of the two Wronskians is the same):
 \be
 \label{28}
 W[a_1(u),..., a_{N+1}(u)] = w_a \ , \qquad
 W[b_1(v),..., b_{N+1}(v) ] = w_b \ , \qquad
 w_a w_b = \tilde{\varepsilon}_{N+1} \ .
 \ee
 The potentials ${\cal U}_n(u)$ ${\cal V}_n(v)$
 can easily be expressed in terms of
 the arbitrary functions $a_i(u)$ and $b_i(v)$, $i=1,...,N$.
 To find the expressions one should differentiate the determinants
 (\ref{28}) to obtain the homogeneous differential equations for
 $a_{N+1}(u)$, $b_{N+1}(v)$. For example, for $N=2$:
 \be
 \label{28a}
 {\cal U}_1(u) = -(a_1 a'''_2 - a'''_1 a_2)/ W[a_1,a_2] , \qquad
 {\cal U}_0(u) = (a'_1 a'''_2 - a'''_1 a'_2)/ W[a_1,a_2] \ .
 \ee

 Let us return to the general solution of Eq.(\ref{21}).
 In fact, considering Eqs.(\ref{28}) as inhomogeneous differential
 equations for $a_{N+1}(u)$, $b_{N+1}(v)$ with arbitrary chosen functions
 $a_i(u)$, $b_i(v)$ ($1\leq i \leq  N$), it is easy to write the explicit
 solution of this problem:
 \be
 \label{31}
 a_{N+1}(u) = \sum_{i=1}^N a_i(u) \int^u_{u_0} d\bar u \ W^{-2}_N(\bar u) \
  M_{N,\,  i}(\bar u) \ .
 \ee
 Here $W_N \equiv W[a_1(u),..., a_N(u)]$  is the Wronskian of
 $N$ arbitrary chosen functions $a_i$ and
 $M_{N, \, i}$ are the complementary minors of the last row in the Wronskian.
 (Replacing $a$ by $b$ and $u$ by $v$ we can find the expression for $b_{N+1}(v)$
 from the same formula (\ref{31})). For the simplest $A_2$-case:
 \bdm
 a_3 (u) = \sum_{i=1}^2 a_i(u) \int^u_{u_0} {{d\bar u} \over {W^2_2(\bar u)}} \
  M_{2,\, i}(\bar u) \equiv  \int^u_{u_0} d\bar u \
  {{a_1(\bar u) a_2(u) - a_1(u) a_2(\bar u)} \over
    {(a_1(\bar u) a_2^{\prime}(\bar u) - a_1^{\prime}(\bar u) a_2(\bar u))^2}} \ .
 \edm
 Thus we have found the expression for the basic solution $X_1$ in terms
 of $2N$ arbitrary chiral functions $a_i(u)$ and $b_i(v)$.
 To complete constructing the solution we should
 derive the expressions for all $X_n$ in terms of $a_i$ and $b_i$.
 This can be done with simple combinatorics that allows one to express $X_n$
 in terms of the $n$-th order minors. For example, it is easy to
 derive the expressions for $X_2$:
  \be
  \label{31a}
 X_2 = \varepsilon_1 \Delta_2 (X_1) =
   \varepsilon_1 \sum_{i<j}  W[a_i(u), a_j(u)] \ W[b_i(v), b_j(v)] \ ,
 \ee
 which is valid for any $N \geq 1$ ($i,j = 1,...,N+1$).
 Note that expressions for all $X_n$
 have a similar separated form
 with higher-order determinants.

 Our simple representation of the $A_N$ Toda solution is completely
 equivalent to
 what one can find in \cite{Leznov} but is more convenient for treating
 some problems. For example, it is useful in discussing asymptotic and
 analytic properties of the solutions of the original physical problems.
 It is especially appropriate for constructing wave-like solutions of
 the Toda system which are similar to the wave solutions of the
 $N$-Liouville model. In fact, quite like the Liouville model,
 the Toda equations
 support the wave-like solutions. To derive them let us first identify
 the moduli space of the Toda solutions. Recalling the $N$-Liouville
 case, we may try to identify the moduli space with the space of the
 potentials ${\cal U}_n(u)$, ${\cal V}_n(v)$. Possibly, this is not
 the best choice and, in fact, in the Liouville case we finally made
 a more useful
 choice suggested by the solution of the constraints.
 For our present purposes the choice of the potentials is as good as any
 other because each choice of ${\cal U}_n(u)$ and ${\cal V}_n(v)$ defines
 some solution and, vice versa, any solution given by the set of the
 functions ($a_1(u),..., a_{N+1} (u)$), ($b_1 (v),..., b_{N+1} (v)$)
 satisfying
 the Wronskian constraints (\ref{28}) defines the corresponding set of
 potentials (${\cal U}_0(u ),...,{\cal U}_{N-1}(u)$),
 (${\cal V}_0(v),...,{\cal V}_{N-1}(v)$).

 Now, as in the Liouville case, we may consider the reduction of the
 moduli space to the space of constant `vectors' $(U_0,...,U_{N-1})$,
 $(V_0,...,V_{N-1})$. The fundamental solutions of the equations (\ref{27})
 with these potentials are exponentials (in the nondegenerate case):
 $\exp (\mu_i u)$, $\exp (\nu_i v)$.
 Then $X_1$ can be written as (for simplicity we take $f_i >0$):
 \be
 \label{32a}
 X_1 = \sum_{i=1}^{N+1} a_i (u) b_i (v)
 = \sum_{i=1}^{N+1} f_i \exp (\mu_i u)\ \exp (\nu_i v)
 \equiv \sum_{i=1}^{N+1} \exp [\mu_i u + u_i)]\ \exp [\nu_i v + v_i)] \ ,
   \ee
 where the parameters must satisfy the
 conditions (\ref{28}). Calculating the determinant $\Delta_{N+1}(X_1)$
 and denoting the standard Vandermonde  determinants by
 \bdm
 D_{\mu} \equiv \prod_{i>j} (\mu_i - \mu_j) \ , \qquad
 D_{\nu} \equiv \prod_{i>j} (\nu_i - \nu_j) \ ,
 \edm
 one can easily find that (\ref{28}) is satisfied if
 \be
 \label{32b}
 \sum_{i=1}^{N+1} \mu_i \, = \, \sum_{i=1}^{N+1} \nu_i = 0 \ ,
  \qquad  \prod_{i=1}^{N+1} f_i \ D_{\mu} \ D_{\nu} =
 \tilde{\varepsilon}_{N+1} \ .
  \ee
  By the way, instead of the last condition we could write the
 equivalent conditions (\ref{28}):
 \be
 \label{32c}
   \prod_{i=1}^{N+1} \exp u_i = w_a \ , \qquad
   \prod_{i=1}^{N+1} \exp v_i = w_b \ , \qquad
   w_a w_b  = (D_{\mu} \ D_{\nu})^{-1} \tilde{\varepsilon}_{N+1}  \ ,
  \ee
 where $\exp u_i$ and $\exp v_i$  are not necessary positive
 (e.g., we can make $\exp u_i $ negative by supposing that $u_i$
 has the imaginary  part $i \pi$) but here we mostly consider
 positive $f_i$.

 In this reduced case we may regard the space of the parameters
 ($\mu_i$, $\nu_i$, $u_i$, $v_i$) as the new moduli space,
 in complete agreement with the Liouville case. Having the basic
 solution $X_1$ given by Eqs.(\ref{32a})-(\ref{32b}) it is not
 difficult to derive $X_n$ recursively by using (\ref{19}).
 For illustration, consider the simplest TL theory $A_1+A_2$.
 Then $X_2$ is given by (\ref{31a}) and (\ref{32a})-(\ref{32b}):
 \be
 \label{32d}
 X_2  = \varepsilon_2 (D_{\mu} \ D_{\nu})^{-1} \sum_{i=1}^3
 \exp [-\mu_i u - u_i)]\ \exp [-\nu_i v - v_i)]
 (\mu_j - \mu_k) (\nu_j - \nu_k) \ ,
 \ee
 where $(ijk)$ is a cyclic permutation of $(123)$.
 The next step is to consider the constraints (\ref{f60}),
 where $X_3$ is the solution of the Liouville equation
 (in order not to mix it with the $X_3$ of the $A_2$-solution
  that is equal to 1, we
 better denote it by $\tilde{X}$).
  Of course, we should suppose that this solution has the form
  (\ref{f45}) with exponential functions (\ref{32a}). In the Liouville
  case $N=2$ and thus $\tilde{X}^{''} / \tilde{X}$ is simply
  $\tilde{\mu}^2$ or $\tilde{\nu}^2$ (see (\ref{f55})).

  Now, using Eqs.(\ref{32a})-(\ref{32d}), one can find that the
  constraints are equivalent to the following equations:
  \be
  \label{36}
  \sum_{i<j} (\mu_i -\mu_k) (\nu_j-\nu_k)[3\mu_k^2 - C_{\mu}] = 0 \ , \qquad
  \sum_{i<j} (\mu_i -\mu_k) (\nu_j-\nu_k)[3\nu_k^2 - C_{\nu}] = 0 \ ,
  \ee
  \be
  \label{37}
  \mu_1^2 + \mu_2^2 + \mu_1 \mu_2  =  C_{\mu} \ , \qquad
  \nu_1^2 + \nu_2^2 + \nu_1 \nu_2  =  C_{\nu} \ ,
  \ee
  where $C_{\mu}$ and $C_{\nu}$ represent contribution of the Liouville
  term. Computing the sums in Eq.(\ref{36}) we find that equations (\ref{36})
  are equivalent to the relations
  \be
  \label{38}
  [(\mu_1^2 + \mu_2^2 + \mu_1 \mu_2) - C_{\mu}] \sum \mu_i \nu_i = 0 \ ,
  \quad
  [(\nu_1^2 + \nu_2^2 + \mu_1 \nu_2) - C_{\nu}] \sum \mu_i \nu_i = 0 \ ,
   \ee
  which are satisfied as soon as Eqs.(\ref{37}) are satisfied.

  It is not difficult to check that the potentials ${\cal U}_1(u)$,
  ${\cal V}_1(u)$ for the exponential solutions are
 \be
 \label{39}
 {\cal U}_1(u) = - (\mu_1^2 + \mu_2^2 + \mu_1 \mu_2) , \qquad
 {\cal V}_1(u) = - (\nu_1^2 + \nu_2^2 + \mu_1 \nu_2) ,
 \ee
 and thus the constraints have extremely simple and natural form:
 \be
 \label{40}
 {\cal U}_1 + C_{\mu} =0 \ , \qquad
 {\cal V}_1 + C_{\nu} = 0.
 \ee

  For the (1+1)-dimensional $A_2$ Toda plus Liouville case
 we have found that the constraints (\ref{f40}) with any number of
 Liouville terms are satisfied for the general solution (i.e. if we put
 into (\ref{40}) the expression (\ref{28a})). Note that in case of just
 one Liouville term this does not help to find an {\bf explicit} solution
 of the constraint. However, if the number of the Liouville terms in
 Eq.(\ref{f40}) is greater than two, and if $\sum \gamma_n$
 for these terms vanishes, one can easily derive the explicit general
 solution by applying the method described in \cite{ATF2},  \cite{ATF5}.
  A detailed account of these results will be published elsewhere.

\section{Conclusion}
 Let us briefly summarize the main results and possible applications.
 We introduced a simple and compact formulation of the general
 (1+1)-dimensional dilaton gravity with multi-exponential potentials
 and derived the conditions allowing to find its explicit solutions
 in terms of the Toda theory. The simplest class of theories
 satisfying these conditions is the Toda-Liouville theory\footnote{
 In \cite{ATF6}, it was shown that the models with the potential
 independent of the dilaton $\f$ can be explicitly solved if $A_{mn}$
 is any Cartan matrix. In this case adding the Liouville part is
 unnecessary.}. We proposed a simple approach to solving the
 equations and constraints in the case of the $A_N$ Toda part.

 Of special interest are simple exponential solutions derived in
 the last section. They explicitly unify the static (black hole)
 solutions\footnote{These solutions normally have two horizons
 defined by zeroes of the metric, i.e. $F \rightarrow -\infty$.
 In the Liouville case they were analyzed in \cite{ATF4},
 \cite{ATF1}, \cite{ATF2}, \cite{ATF5}. },
 cosmological models, and waves of the Toda matter coupled
 to gravity. Some of these solutions can be related to cosmologies
 with  spherical inhomogeneities or to evolving black holes but
 this requires special studies.
 Earlier we studied similar but simpler solutions in the
 $N$-Liouville theories in paper \cite{ATF5}. The main results of that
 paper, in particular, the existence of nonsingular exponential solutions,
 are true also in the Toda-Liouville theory.

 Note that one-dimensional Toda-Liouville cosmological models were
 met long time ago in dimensional reductions of higher-dimensional
 (super)gravity theories (see, e.g., \cite{Pope}). Considerations of the
 two-dimensional TL theories of this paper are equally applicable to
 the one-dimensional case. A preliminary discussion can be found in
 \cite{ATF6} and the detailed consideration will be published elsewhere,
 together with a detailed presentation of the results that were only
 briefly described here.

 Finally, note that here we only give an account of
 the first part of the report presented at the workshop `Quarks-2008'
 (see the presentation of our report at the site http://inr.ac.ru).
 In the second part, a brief summary and a new interpretation
 of A.Einstein's paper \cite{1923} was proposed by one of the
 present authors (ATF). The proposal is that Einstein's theory
 (that he regarded as a unified theory of gravity and electromagnetism)
 is in fact a first unified model of dark energy (dictated by the
 geometry cosmological constant) and dark matter
 (dictated by the geometry neutral massive vector field
 coupled to gravity only). Unfortunately even one-dimensional spherically
 symmetric reductions of this theory are not integrable
 (a preliminary analysis\footnote{It is interesting that the static solutions
 may have two horizons, like the Reissner - Nordstroem black holes, although there is
 no electric charge in the model. }
 of these solutions can be found in
 http://atfilippov.googlepages.com/ogiev.ppt).

\newpage


 {\bf Acknowledgment:}

 One of the authors appreciates financial support from
 the Department of Theoretical Physics of the University of Turin and INFN
(Turin Section), where this work was completed.

 This work was supported in part by the Russian Foundation for Basic
 Research (Grant No. 06-01-00627-a).


\end{document}